\documentclass[onecolumn,11pt]{IEEEtran}
\usepackage{url,epsfig}
\usepackage{graphicx}
\usepackage{amsmath, amsthm, amssymb}
\usepackage{multirow}
\usepackage{amsfonts}
\usepackage{graphicx}
\usepackage{setspace}
\usepackage{datetime}
\usepackage{multirow}
\usepackage{color}

\usepackage[square, comma, sort&compress]{natbib}
\newcommand{\bfm}[1]{\mbox{\boldmath $#1$}}
\def\thanks#1{
    \protected@xdef\@thanks{\@thanks
        \protect\footnotetext[\the\c@footnote]{#1}}%
}

\newtheorem{thm}{Theorem}

\newtheorem{lem}{Lemma}
\newtheorem{prop}{Proposition}
\newtheorem{corol}{Corollary}
\date{\today}
\begin{document}
\title{Achievable Rates for Noisy Channels with Synchronization Errors\footnote{This research is funded by the National Science
Foundation under contract NSF-TF 0830611.}\footnote{Part of this work is submitted to 2012 IEEE International Symposium on Information Theory (ISIT).}}
\author{\IEEEauthorblockN{Mojtaba Rahmati and Tolga M. Duman}\\
\IEEEauthorblockA{School of  Electrical, Computer and Energy Engineering, Fulton Schools of Engineering\\
Arizona State University, Tempe, AZ 85287--5706, USA\\
Email: mojtaba@asu.edu} and duman@asu.edu}
\maketitle

\begin{abstract}
We develop several lower bounds on the capacity of binary input symmetric output channels with synchronization errors which also suffer from other types of impairments such as substitutions, erasures, additive white Gaussian noise (AWGN) etc. More precisely, we show that if the channel with synchronization errors can be decomposed into a cascade of two channels where only the first one suffers from synchronization errors and the second one is a memoryless channel, a lower bound on the capacity of the original channel in terms of the capacity of the synchronization error-only channel can be derived. To accomplish this, we derive lower bounds on the mutual information rate between the transmitted and received sequences (for the original channel) for an arbitrary input distribution, and then relate this result to the channel capacity. The results apply without the knowledge of the exact capacity achieving input distributions. A primary application of our results is that we can employ any lower bound derived on the capacity of the first channel (synchronization error channel in the decomposition) to find lower bounds on  the capacity of the (original) noisy channel with synchronization errors. We apply the general ideas to several specific classes of channels such as synchronization error channels with erasures and substitutions, with symmetric $q$-ary outputs and with AWGN explicitly, and obtain easy-to-compute bounds. We illustrate that, with our approach, it is possible to derive tighter capacity lower bounds compared to the currently available bounds in the literature for certain classes of channels, e.g., deletion/substitution channels and deletion/AWGN channels (for certain signal to noise ratio (SNR) ranges).
\end{abstract}

\noindent

\vspace{1\baselineskip}

\doublespacing
\section{Introduction}
Depending on the transmitting medium and the particular design, different limiting factors degrade the performance of a general communication system. For instance, imperfect alignment of the transmitter and receiver clocks may be one such factor resulting in a synchronization error channel modeled typically through insertion and/or deletion of symbols. Other factors include the effects of additive noise at the receiver etc. The main objective of this paper is to study the combined effects of the synchronization errors and additive noise type impairments and in particular to ``decouple'' the effects of the synchronization errors from the other parameters and obtain expressions relating the channel capacity of the combined model and the synchronization error-only channel. 

We focus on achievable rates for channels which can be considered as a concatenation of two independent channels where the first one is a binary channel suffering only from synchronization errors and the second one is either a memoryless binary input symmetric $q$-ary output channel (BSQC) or a binary input AWGN (BI-AWGN) channel. For instance, the first channel can be a binary insertion/deletion channel and the second one can be a binary symmetric channel (BSC) or a substitution/erasure channel (a ternary output channel $q=3$). Our development starts with the ternary ($q=3$) and quaternary ($q=4$) output cases, respectively, then we generalize the results to a general $q$-ary output case. Specifically, we obtain achievable rates for the concatenated channel in terms of  the capacity of the synchronization error channel by lower bounding the information rate of the concatenated channel for input distributions which achieve the capacity of the synchronization error-only channel and the parameters of the memoryless channel. The lower bounds are derived without the use of the exact capacity achieving input distribution of the synchronization error channel, hence any existing lower bound on the capacity (of the synchronization error-only channel) can be employed to obtain an achievable rate characterization for the original channel model of interest. 

By channels with synchronization errors we refer to the binary memoryless channels with synchronization errors as described by Dobrushin in~\cite{dobrushin} where every transmitted bit is independently replaced with a random number of symbols (possibly empty string, i.e. a deletion event is also allowed), and the transmitter and receiver have no information about the position and/or the pattern of the insertions/deletions. Different specific models on channels with synchronization errors are considered in the literature. Insertion/deletion channels are used as common models for channels with synchronization errors, e.g., the Gallager insertion/deletion channel~\cite{gallager}, the sticky channel~\cite{sticky} and the segmented insertion/deletion channel~\cite{liu2010segmented}. 

Dobrushin~\cite{dobrushin} proved that Shannon's theorem applies for a memoryless channel with synchronization errors by demonstrating that information stability holds for memoryless channels with synchronization errors. That is, for the capacity of the synchronization error channel, $C_s$ we can write $\displaystyle C_s=\lim_{N\to \infty}\displaystyle
\max_{P\left(\bfm{X}\right)}\dfrac{1}{N}I(\bfm{X};\bfm{Y})$, where $\bfm X$ and $\bfm Y$ are the transmitted and received sequences, respectively, and $N$ is the length of the transmitted sequence. Therefore, the information and transmission capacities of the memoryless channels with synchronization errors are equal and we can employ any lower bound on the information capacity as a lower bound on the transmission capacity of a channel with synchronization errors.

There are many papers deriving upper and/or lower bounds on the capacity of the insertion/deletion channels, e.g., see~\cite{drinea2007improved,drinea2007,dario,kanoria,asymptotic,IT-paper}; however, only a very few results exist for insertion/deletion channels with substitution errors, e.g.~\cite{gallager,dario2} or in the presence of AWGN, e.g.~\cite{zeng2005bounds,junhu}. Our interest is on the latter, in fact, on more general models incorporating erasures as well as $q$-ary channel outputs. 

Let us review some of the existing relevant results on insertion/deletion channels in a bit more detail. In~\cite{gallager}, Gallager considers a channel model with substitution and insertion/deletion errors (sub/ins/del) where each bit gets deleted with probability $p_d$, replaced by two random bits with probability $p_i$, correctly received with probability 
$p_c=(1-p_d-p_i)(1-p_s)$, and changed with probability $p_f=(1-p_d-p_i)p_s$, and derives a lower bound on the channel capacity (in bits/use) given by 
\begin{equation}  \label{eq:LB-gallager}
C \geq 1+p_d \log{p_d}+p_i \log{p_i}+p_c
\log{p_c}+p_f\log(p_f),
\end{equation}
where $\log(.)$ denotes logarithm in base~2. Fertonani and Duman in~\cite{dario2} develop several upper and lower bounds on the capacity of the ins/del/sub channel, where they employ a genie-aided decoder that is supplied with side information about some suitably selected random processes. Therefore, an auxiliary memoryless channel is obtained in such a way that the Blahut-Arimoto algorithm (BAA) can be employed to obtain upper bounds on the capacity of the original channel. Furthermore, it is shown that by subtracting some quantity from the derived upper bounds which is, roughly speaking, more than extra information provided by the side information, lower bounds on the capacity can also be derived. In~\cite{junhu}, Monte Carlo simulation based results are used to estimate information rates of different insertion and/or deletion channels in the absence or presence of intersymbol interference (ISI) in addition to AWGN with independent uniformly distributed (i.u.d.) input sequences. In~\cite{zeng2005bounds}, the synchronization errors are modeled as a Markov process and simulation results are used to compute achievable information rates of an ISI channel with synchronization errors in the presence of AWGN. In~\cite{IT-paper}, Rahmati and Duman compute analytical lower bounds on the capacity of the i.i.d. del/sub and i.i.d. del/AWGN channels, by lower bounding the mutual information rate between the transmitted and received sequences for i.u.d. input sequences focusing on small deletion probabilities. 

The paper is organized as follows. In Section~\ref{sec-channel_model}, we formally give the models for binary input symmetric $q$-ary output channels with synchronization errors and BI-AWGN channels with synchronization errors. In~\ref{sec-main_lemmas}, we give two lemmas and one proposition which will be useful in the proof of the result on BSQC channels with synchronization errors. In Section~\ref{sec-BSQC}, we initially focus on a substitution/erasure/synchronization error channel (abbreviated as sub/ers/synch channel) which is a binary input symmetric ternary output channel, and then on a binary input symmetric quaternary output channel. After that we extend the results to the case of more general symmetric $q$-ary output channels. In Section~\ref{sec-AWGN}, we lower bound the capacity of a synchronization error channel with AWGN (abbreviated as AWGN/synch channel) in terms of the capacity of the underlying synchronization error only channel. More precisely, we generalize the results on BSQC channels with synchronization errors when $q$ goes to infinity. We present several numerical examples illustrating the derived results in~Section~\ref{sec-num}. Finally, we conclude the paper in Section~\ref{sec-conc}.\color{black}

\section{Channel Models}\label{sec-channel_model}
A general memoryless channel with synchronization errors~\cite{dobrushin} is defined via a stochastic matrix $\{p(y_i|x_i),y_i \in {\cal{Y}},{x_i} \in {\cal{X}}\}$ where $\cal{X}$ is the input alphabet (e.g., for a binary input channel ${\cal{X}}=\{0,1\}$), and $\cal{Y}$ is (possibly empty) the set of output symbols, $0\leq p(y_i|x_i)\leq 1$, and $\sum_{y_i\in\cal Y}p(y_i|x)=1$. 
%
%
As a particular instance of this channel, if $p(y_i=\emptyset |x_i)=p_d$ ($\emptyset$ denoting the null string) and $p(y_i=x_i)=1-p_i$, we obtain an i.i.d. deletion channel. 

\subsection{Binary Input Symmetric $q$-ary Output Channel with Synchronization Errors}\label{sec_ch_mod_q}
\begin{figure}
    \begin{center}
    \includegraphics[trim=35mm 100mm 50mm 68mm,clip,  width=.6\textwidth]{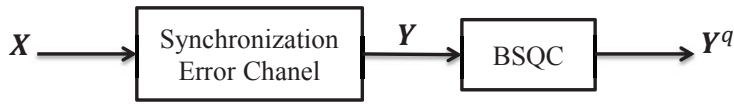}
    \caption{Binary input symmetric $q$-ary output channel with synchronization errors.}
    \label{fig:synch-BSQC}
    \end{center}
\end{figure}

\begin{figure}[t]
    \begin{center}
    \includegraphics[trim=30mm 109mm 25mm 20mm,clip,  width=.65\textwidth]{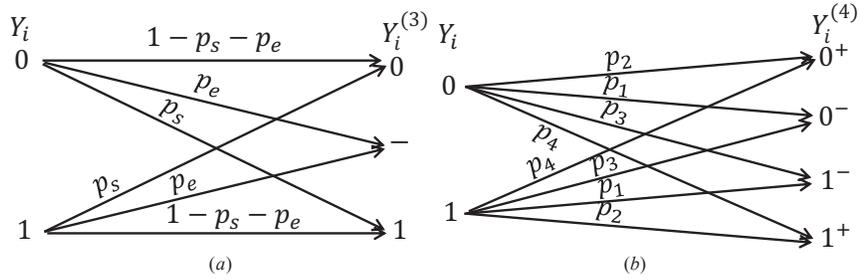}\vspace*{-.1in}
    \caption{(a) Input-output relation in the substitution/erasure channel ($P(Y^{(3)}_i|Y_i)$ for all $1\leq i\leq |\bfm y|$). (b) Input-output relation in the binary input quaternary output channel ($P(Y^{(4)}_i|Y_i)$ for all $1\leq i\leq |\bfm y|$).}
    \label{fig:3_4_ary}
    \end{center}
    \vspace*{-.25in}
\end{figure}

By a binary input symmetric $q$-ary output channel (BSQC) with synchronization errors, we refer to a channel which can be considered as a concatenation of two independent channels, depicted in Fig~\ref{fig:synch-BSQC}, such that the first one is a channel with only synchronization errors with input sequence $\bfm X$ and output sequence $\bfm Y$, and the second one is a BSQC with input sequence $\bfm Y$ and output sequence $\bfm Y^{(q)}$, where by a symmetric channel we refer to the definition given in~\cite[p.~94]{gallager_book}. In other words, a channel is symmetric if by dividing the columns of the transition matrix into sub-matrices, in each sub-matrix, each row is a permutation of any other row and each column is a permutation of any other column. For example, a channel with independent substitution, erasure and synchronization errors (sub/ers/synch channel) can be considered as a concatenation of a channel with only synchronization errors with input sequence $\bfm X$ and output sequence $\bfm Y$ and a substitution/erasure channel (binary input ternary output channel) with input sequence $\bfm Y$ and output sequence $\bfm Y^{(3)}$. In a  substitution/erasure channel, each bit is independently flipped with probability $p_s$ or erased with probability $p_e$, as illustrated in Fig.~\ref{fig:3_4_ary}. (a).
Another example is a binary input symmetric quaternary output channel with synchronization errors which can be decomposed into two independent channels such that the first one is a memoryless synchronization error channel and the second one is a memoryless binary input symmetric quaternary output channel 
shown in Fig.~\ref{fig:3_4_ary}. (b).

%
\subsection{BI-AWGN Channels with Synchronization Errors}\label{sec-AWGN_model}
\begin{figure}
    \begin{center}
    \includegraphics[trim=35mm 100mm 50mm 68mm,clip,  width=.6\textwidth]{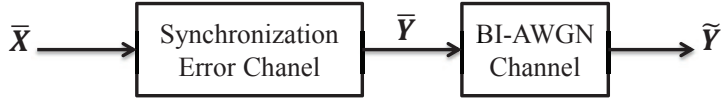}
    \caption{AWGN channel with synchronization errors.}
    \label{fig:synch-AWGN}
    \end{center}
\end{figure}

In a BI-AWGN channel with synchronization errors, bits are transmitted using binary phase shift keying (BPSK) and the received signal contains AWGN in addition to the synchronization errors. As illustrated in Fig. \ref{fig:synch-AWGN}, this channel can be considered as a cascade of two independent channels where the first one is a synchronization error channel and the second one is a BI-AWGN channel. We use $\bar{\bfm X}$ to denote the input sequence to the first channel which is a BPSK modulated version of the binary input sequence $\bfm X$, i.e., $\bar{X}_i=1-2X_i$ and $\bar{\bfm Y}$ to denote the output sequence of the first channel and input to the second one. $\widetilde{\bfm Y}$ is the output sequence of the second channel that is the noisy version of $\bar{\bfm Y}$, i.e.,
\begin{equation}
\widetilde{Y}_i^d= \bar{Y}_i^d + Z_i,\nonumber
\end{equation}
where $Z_i$'s are i.i.d. zero mean Gaussian random variables with variance $\sigma^2$, and $\widetilde{Y}_i^d$ and
$\bar{Y}_i^d$ are the $i^{th}$ received and transmitted bits of the second channel, respectively.

  \subsection{Simple Example of a Synchronization Error Channel Decomposition into Two Independent Channels}
  The procedure in finding the capacity bounds used in this paper can be employed for any channel which can be decomposed into two independent channels such that the first one is a memoryless synchronization error channel and the second one is a symmetric memoryless channel with no effect on the length of the input sequence. Therefore, if we can decompose a given synchronization error channel into two channels with described properties, we can derive lower bounds on the capacity of the synchronization error channel. The advantage of this decomposition is that decomposing the original synchronization error channel into a well characterized synchronization error channel and a memoryless channel could be done in such a way that lower bounding the capacity of the new synchronization error channel be simpler than the capacity analysis of the original synchronization error channel. In Table~\ref{tb_ex}, we provide an example of a hypothetical channel with synchronization errors that can be decomposed into a different synchronization error channel and a memoryless binary symmetric channel (BSC). In Table~\ref{tb_decompose}, the two channels used in the decomposition are given.
 \begin{table}[t]
  \centering
  \caption{Transition probabilities of the hypothetical synchronization error channel.}
  \label{tb_ex}
  \begin{tabular}{|c|c|c|c|c|c|c|}
  \hline
  \multicolumn{1}{|c|}{} &
  \multicolumn{6}{|c|}{$P(Y_j | X_j)$}\\
  \hline
  $X_j$ & $Y_j=0$ & $Y_j=1$ & $Y_j=00$ & $Y_j=01$ & $Y_j=10$ &$Y_j=11$\\
  \hline
  0 & $\left(\frac{1}{2}-(\alpha+\beta)\right)\left(1+\sqrt{\frac{\alpha-\beta}{\alpha+\beta}}\right) $ & $\left(\frac{1}{2}-(\alpha+\beta)\right)\left(1-\sqrt{\frac{\alpha-\beta}{\alpha+\beta}}\right) $& $\alpha $ & $\beta$ & $\beta$ & $\alpha$\\
  \hline
  1 & $\left(\frac{1}{2}-(\alpha+\beta)\right)\left(1-\sqrt{\frac{\alpha-\beta}{\alpha+\beta}}\right) $ & $\left(\frac{1}{2}-(\alpha+\beta)\right)\left(1+\sqrt{\frac{\alpha-\beta}{\alpha+\beta}}\right) $& $\alpha $ & $\beta$ & $\beta$ & $\alpha$\\
  \hline
  \end{tabular}
  \end{table}

\begin{table}[t]
\caption{Transition probabilities of two independent channels giving rise to the synchronization error channel given in Table ~\ref{tb_ex}.}
\label{tb_decompose}
\begin{minipage}[b]{0.55\linewidth}\centering
\begin{tabular}{|c|c|c|c|c|}
\hline
  \multicolumn{1}{|c|}{} &
  \multicolumn{4}{|c|}{$P(Z_j | X_j)$}\\
  \hline
  $X_j$ & $Z_j=0$ & $Z_j=1$ & $Z_j=00$ & $Z_j=11$\\
  \hline
   0    & $1-2(\alpha+\beta)$ & 0 & $\alpha+\beta$ & $\alpha+\beta$\\
   \hline
   1    & 0 & $1-2(\alpha+\beta)$ & $\alpha+\beta$ & $\alpha+\beta$\\
   \hline
        \end{tabular}
\end{minipage}
\hspace{0.5cm}
\begin{minipage}[b]{0.4\linewidth}
\centering
\begin{tabular}{|c|c|c|}
\hline
  \multicolumn{1}{|c|}{} &
  \multicolumn{2}{|c|}{$P(Y_j | Z_j)$}\\
  \hline
  $Z_j$ & $Y_j=0$ & $Y_j=1$ \\
  \hline
   0    & $0.5+0.5\sqrt{\frac{\alpha-\beta}{\alpha+\beta}}$&$0.5-0.5\sqrt{\frac{\alpha-\beta}{\alpha+\beta}}$\\
   \hline
   1    & $0.5-0.5\sqrt{\frac{\alpha-\beta}{\alpha+\beta}}$&$0.5+0.5\sqrt{\frac{\alpha-\beta}{\alpha+\beta}}$\\
   \hline
        \end{tabular}
\end{minipage}
\end{table}

\section{Entropy Bounds for Binary Input $q$-ary Output Channels with Synchronization Errors}\label{sec-main_lemmas}
In the following two lemmas, we provide a lower bound on the output entropy and an upper bound on the conditional output entropy of the binary input $q$-ary output channel in terms of the the corresponding output entropies of the synchronization error channel, respectively. We then give a proposition that will be useful in the proof of the result on BSQC channels with synchronization errors (note that the following two lemmas hold for any binary input $q$-ary output channels with synchronization errors regardless of any symmetry).

\begin{lem}\label{lem_Hy^{(q)}HY}
In any binary input $q$-ary output channel with synchronization errors and for all non-negative integer values of $q$, we have
\begin{equation}\label{eq_Hy^{(q)}HY}
H(\bfm Y^{(q)})\geq H(\bfm Y)-E_{\bfm M}\left\{\log\left(\sum_{\bfm y^{(q)}}\sum_{\bfm y, p(\bfm y)\neq 0}p(\bfm y^{(q)}|\bfm y,\bfm M)p(\bfm y^{(q)}|\bfm M)\right)\right\},
\end{equation}
where $\bfm M$ is the random variable denoting the length of the received sequence, $\bfm Y$ denotes the output sequence of the synchronization error channel and the input sequence of the binary input $q$-ary output channel, and $\bfm Y^{(q)}$ denotes the output sequence of the binary input $q$-ary output channel.
\end{lem}
\begin{IEEEproof}
By using two different expansions of $H(\bfm Y^{(q)},\bfm M)$, we have
\begin{eqnarray}
H(\bfm Y^{(q)},\bfm M)&=&H(\bfm Y^{(q)})+H(\bfm M|\bfm Y^{(q)})\nonumber\\
&=&H(\bfm Y^{(q)}|\bfm M)+H(\bfm M).
\end{eqnarray}
Hence, we can write
\begin{equation}\label{}
H(\bfm Y^{(q)})=H(\bfm Y^{(q)}|\bfm M)+H(\bfm M),
\end{equation}
where we used the fact that by knowing $\bfm Y^{(q)}$, random variable $\bfm M$ is also known, i.e. $H(\bfm M|\bfm Y^{(q)})=0$. By using the same approach for $H(\bfm Y)$, we have
\begin{equation}
H(\bfm Y)=H(\bfm Y|\bfm M)+H(\bfm M).
\end{equation}
Finally, we can write
\begin{eqnarray}\label{eq_Hy^{(q)}-H}
H(\bfm Y^{(q)})-H(\bfm Y)&=&H(\bfm Y^{(q)}|\bfm M)-H(\bfm Y|\bfm M)\nonumber\\
&=&\sum_{m}p(m)\left[H(\bfm Y^{(q)}|\bfm M=m)-H(\bfm Y|\bfm M=m)\right],
\end{eqnarray}
where $p(m)=P(\bfm M=m)$. On the other hand, due to the definition of the entropy, we can write
\begin{align}\label{eq_Hy^{(q)}-Hm}
H(\bfm Y^{(q)}|\bfm M=m)-H(\bfm Y|\bfm M=m)&=E_{\bfm Y^{(q)}}\{-\log(p(\bfm Y^{(q)}))|\bfm M=m\}-E_{\bfm Y}\{-\log(p(\bfm Y))|\bfm M=m\}\nonumber\\
&=E_{(\bfm Y,\bfm Y^{(q)})}\left\{-\log\left(\frac{p(\bfm Y^{(q)})}{p(\bfm Y)}\right)\bigg|\bfm M=m\right\}\nonumber\\
&=-\sum_{\bfm y^{(q)}}\sum_{\bfm y,p(\bfm y)\neq 0}p(\bfm y^{(q)}|\bfm y,\bfm M=m)p(\bfm y|\bfm M=m)\log\left(\frac{p(\bfm y^{(q)}|\bfm M=m)}{p(\bfm y|\bfm M=m)}\right),\nonumber
\end{align}
where $E_{\bfm Z}\{.\}$ denotes the expected value with respect to the random variable $\bfm Z$. Now due to the fact that $-\log(x)$ is a convex function of $x$, we apply Jensen's inequality to write
\begin{eqnarray}\label{eq_Ee}
H(\bfm Y^{(q)}|\bfm M=m)-H(\bfm Y|\bfm M=m)&\geq & -\log\left(\sum_{\bfm y^{(q)}}\sum_{\bfm y, p(\bfm y)\neq 0}p(\bfm y^{(q)}|\bfm y,\bfm M=m)p(\bfm y|\bfm M=m)\frac{p(\bfm y^{(q)}|\bfm M=m)}{p(\bfm y|\bfm M=m)}\right)\nonumber\\
&=&-\log\left(\sum_{\bfm y^{(q)}}\sum_{\bfm y, p(\bfm y)\neq 0}p(\bfm y^{(q)}|\bfm y,\bfm M=m)p(\bfm y^{(q)}|\bfm M=m)\right).
\end{eqnarray}
By substituting this result into~\eqref{eq_Hy^{(q)}-H}, the proof follows. 
\end{IEEEproof}

\begin{lem}\label{lem_Hy^{(q)}Y|X}
In any binary input $q$-ary output channel with synchronization errors and for any input distribution, we have
\begin{equation}\label{eq_H(y^{(q)}|X)}
H({\bfm Y^{(q)}}|\bfm X)\leq H(\bfm Y|\bfm X)+E\{\bfm M\}H(Y^{(q)}_j|Y_j),
\end{equation}
where $Y_j$ denotes the $j$-th output bit of the synchronization error channel and $j$-th input bit of the binary input $q$-ary output channel and $Y^{(q)}_j$ denotes the output symbol of the binary input $q$-ary output channel corresponding to the input bit $Y_j$.
\end{lem}
\begin{IEEEproof}
For the conditional output entropy, we can write
\begin{eqnarray}\label{eq_H(y^{(q)}x)}
H(\bfm Y^{(q)},\bfm Y|\bfm X)&=&H(\bfm Y^{(q)}|\bfm X)+H(\bfm Y|\bfm Y^{(q)},\bfm X)\nonumber\\
&=&H(\bfm Y|\bfm X)+H(\bfm Y^{(q)}|\bfm Y,\bfm X)\nonumber\\
&=&H(\bfm Y|\bfm X)+H(\bfm Y^{(q)}|\bfm Y),
\end{eqnarray}
where the last equality follows since $\bfm X\to \bfm Y\to\bfm Y^{(q)}$ form a Markov chain.
Therefore,
\begin{eqnarray}
H(\bfm Y^{(q)}|\bfm X)&=&H(\bfm Y|\bfm X)+H(\bfm Y^{(q)}|\bfm Y)-H(\bfm Y|\bfm X,\bfm Y^{(q)})\nonumber\\
&\leq&H(\bfm Y|\bfm X)+H(\bfm Y^{(q)}|\bfm Y).
\end{eqnarray}
On the other hand, by using the fact that by knowing $\bfm Y$, $\bfm M$ is also known, we have
\begin{equation}
H(\bfm Y^{(q)}|\bfm Y)=H(\bfm Y^{(q)}|\bfm M,\bfm Y).
\end{equation}
Furthermore, since the second channel is memoryless, we obtain
\begin{eqnarray}
H(\bfm Y^{(q)}|\bfm Y,\bfm M)&=&\sum_m p(m)H(\bfm Y^{(q)}|\bfm Y,\bfm M=m)\nonumber\\
&=&\sum_m p(m)m H(Y^{(q)}_j|Y_j)\nonumber\\
&=&E_{\bfm M}\left\{M\right\}H(Y^{(q)}_j|Y_j),
\end{eqnarray}
which concludes the proof.
\end{IEEEproof}

By combining the results of Lemmas~\ref{lem_Hy^{(q)}HY} and \ref{lem_Hy^{(q)}Y|X}, we obtain 
\begin{equation}\label{eq_IqI}
I(\bfm X;\bfm Y^q)\geq I(\bfm X;\bfm Y)-E_{\bfm M}\left\{\log\left(\sum_{\bfm y^{(q)}}\sum_{\bfm y, p(\bfm y)\neq 0}p(\bfm y^{(q)}|\bfm y,\bfm M)p(\bfm y^{(q)}|\bfm M)\right)\right\}-E\{\bfm M\}H(Y^{(q)}_j|Y_j),
\end{equation}
which gives a lower bound on the mutual information between the transmitted and received sequences of the concatenated channel $I(\bfm X;\bfm Y^q)$ in terms of the mutual information between the transmitted and received sequences of the synchronization error channel $I(\bfm X;\bfm Y)$.

\begin{prop}\label{prop_I-C}
For any $\bfm X$, $\bfm Y$ and $\bfm Y^{(q)}$ forming a Markov chain $\bfm X\to\bfm Y\to\bfm Y^{(q)}$, if 
\begin{equation}
I(\bfm X;\bfm Y^{(q)})\geq I(\bfm X;\bfm Y)+A,\nonumber
\end{equation} 
where $A$ is a constant, then the capacity of the channels  $\bfm X\to \bfm Y^{(q)}$ ($C_{\bfm X\to \bfm Y^{(q)}}$) and $\bfm X\to\bfm Y$ ($C_{\bfm X\to\bfm Y}$) satisfy
\begin{equation}
C_{\bfm X\to \bfm Y^{(q)}}\geq C_{\bfm X\to\bfm Y}+A.
\end{equation} 

\end{prop}
\begin{IEEEproof}
Using the input distribution which achieves the capacity of the channel $\bfm X\to\bfm Y$, $P(\bfm X)$, we can write
\begin{eqnarray}
\lim_{n\to\infty}\frac{1}{n}I(\bfm X;\bfm Y^{(q)}(\bfm X))&\geq& \lim_{n\to\infty}\frac{1}{n}I(\bfm X;\bfm Y(\bfm X))+A\nonumber\\
&=&C_{\bfm X\to \bfm Y}+A.
\end{eqnarray}
Hence, for the capacity of the channel $\bfm X\to \bfm Y^{(q)}$, we have
\begin{eqnarray}
C_{\bfm X\to \bfm Y^{(q)}}&=&\lim_{n\to\infty}\frac{1}{n}\max_{P(\bfm X)}I(\bfm X;\bfm Y^{(q)})\nonumber\\
&\geq&\lim_{n\to\infty}\frac{1}{n}I(\bfm X;\bfm 
Y^{(q)}(X))\nonumber\\
&\geq& C_{\bfm X\to \bfm Y}+A.
\end{eqnarray}
\end{IEEEproof}

Due to the result in~\eqref{eq_IqI} and the result of Proposition~\ref{prop_I-C}, the capacity of the concatenated channel can be lower bounded in terms of the capacity of the synchronization error channel and the parameters of the second (memoryless) channel.
\section{Achievable Rates over Binary Input Symmetric $q$-ary Output Channels with Synchronization Errors}\label{sec-BSQC}
In this section, we focus on BSQC channels with synchronization errors (as introduced in Section~\ref{sec_ch_mod_q}) and provide lower bounds on their capacity. We first develop the results for sub/ers/synch channel and binary input symmetric quaternary output channel, respectively. Then give the results for general (odd and even) $q$, respectively.

\subsection{Substitution/Erasure Channels with Synchronization Errors}
The following theorem gives a lower bound on the capacity of the sub/ers/synch channel with respect to the capacity of the synchronization error channel. In a sub/ers channel, every transmitted bit is either flipped with probability of $p_s$, or erased with probability of $p_e$ or received correctly with probability of $1-p_s-p_e$ independent of each other.
\begin{thm}\label{thm-synch-sub-ers}
The capacity of the sub/ers/synch channel $C_{ses}$ can be lower bounded by
\begin{equation}\label{LB_synch-sub-ers}
C_{ses}\geq C_{s}-r\left[H(p_s,p_e,1-p_s-p_e)+\log\left((1-p_e)^2+2p_e^2\right)\right],
\end{equation}
where $C_{s}$ denotes the capacity of the synchronization error channel, $r=\lim_{n\to\infty}\frac{E\left\{\bfm M\right\}}{n}$, $n$ and $m$ denote the length of the transmitted and received sequences, respectively.
\end{thm}


Before giving the proof of Theorem~\ref{thm-synch-sub-ers}, we consider some special cases of this result. Since we have considered the general synchronization error channel model of Dobrushin~\cite{dobrushin}, the lower bound~\eqref{LB_synch-sub-ers} holds for many different models on channels with synchronization errors. A popular model for channels with synchronization errors is the Gallager's ins/del model\footnote{In fact, Gallager's model in general refers to a channel with insertion, deletion and substitution errors, but with Gallager's ins/del model we refer to the case with $p_s=0$ (i.e., substitution error probability being zero).} in which every transmitted bit is either deleted with probability of $p_d$ or replaced with two random bits with probability of $p_i$ or received correctly with probability of $1-p_d-p_i$ independent of each other while neither the transmitter nor the receiver have any information about the insertion and/or deletion errors. If we employ the Gallager's model in deriving the lower bounds, for the parameter $r$, we have
\begin{eqnarray}\label{p(d,i)}
r&=&\lim_{n\to\infty}\frac{E\{\bfm M\}}{n}\nonumber\\
&=&\lim_{n\to\infty}\frac{1}{n} n E\{|s_j|\}\nonumber\\
&=&1-p_d+p_i,
\end{eqnarray}
where $|s_j|$ denotes the length of the output sequence in one use of the ins/del channel, and the equality results since the channel is memoryless.
By utilizing the result of~\eqref{p(d,i)} in~\eqref{LB_synch-sub-ers}, we obtain the following two corollaries.
\begin{corol}
The capacity of the sub/ers/ins/del channel $C_{seid}$ is lower bounded by
\begin{equation}\label{LB_ins-del-sub-ers}
C_{seid}\geq C_{id}-(1-p_d+p_i)\left[H(p_s,p_e,1-p_s-p_e)+\log\left((1+p_e)^2+2p_e^2\right)\right],
\end{equation}
where $C_{id}$ denotes the capacity of an insertion/deletion channel with parameters $p_d$ and $p_i$.
\end{corol}
Taking $p_e=0$ in this channel model gives the ins/del/sub channel, hence we have the following corollary.
\begin{corol}
The capacity of the ins/del/sub channel $C_{ids}$ can be lower bounded by
\begin{equation}\label{LB_ins-del-sub}
C_{ids}\geq C_{id}-(1-p_d+p_i)H_b(p_s),
\end{equation}
\end{corol}


To prove Theorem~\ref{thm-synch-sub-ers}, we need the following two lemmas. 
In the first one we give a lower bound on the output entropy of the sub/ers/synch channel related to the output entropy of the insertion/deletion channel, while in the second one we give an upper bound on the conditional output entropy of the sub/ers/synch channel, related to the conditional output entropy of the insertion/deletion channel.

\begin{lem}\label{lem_Hy^{(3)}HYe}
For a sub/ers/synch channel, for any input distribution, we have
\begin{equation}
H(\bfm Y^{(3)})\geq H(\bfm Y)-E\{\bfm M\}\log\left((1-p_e)^2+2p_e^2\right),
\end{equation}
where $\bfm Y$ denotes the output sequence of the synchronization error channel and input sequence of the substitution/erasure channel, and $\bfm Y^{(3)}$ denotes the output sequence of the substitution/erasure channel.
\end{lem}
\begin{IEEEproof}
Using the result of Lemma~\ref{lem_Hy^{(q)}HY}, we only need to obtain an upper bound on $$\displaystyle \sum_{\bfm y^{(3)}}\sum_{\bfm y,p(\bfm y)\neq 0}p(\bfm y^{(3)}|\bfm y,\bfm M=m)p(\bfm y^{(3)}|\bfm M=m)$$ for all values of $m$.
On the other hand for $p(\bfm y^{(3)}|\bfm y,\bfm M=m)$, we have
\begin{eqnarray}
p(\bfm y^{(3)}|\bfm y,\bfm M=m)&=&\prod_{i=1}^{m} p(Y^{(3)}_i|Y_i)\nonumber\\
&=&p_e^{j_1}p_s^{j_2}(1-p_s-p_e)^{m-j_1-j_2},
\end{eqnarray}
where $j_1$ denotes the number of transitions $0\to -$ or $1\to -$ and $j_2$ denotes the number of transitions $0\to 1$ or $1\to 0$. E.g., $p(011-|0000)=p(0|0)p(1|0)p(1|0)p(-|0)=p_ep_s^2(1-p_e-p_s)$. On the other hand, for a fixed output sequence $\bfm y^{(3)}$ of length $m$ with $j_1$ erased symbols $``-"$, there are $2^{j_1}{m-j_1\choose {j_2}}$ possibilities among all $m$-tuples such that $d(\bfm y^{(3)})_{e}=j_1$, i.e., the number of erased symbols in $\bfm y^{(3)}$, and $d(\bfm y,\bfm y^{(3)})_{s}=j_2$, i.e., the number of positions in $\bfm y$ and $\bfm y^{(3)}$ in which $Y^{(3)}_j$'s are the flipped versions of $Y_j$, therefore we can write
\begin{eqnarray}\label{eq_py0}
\sum_{\bfm y,p(\bfm y)\neq 0}p(\bfm y^{(3)}|\bfm y,\bfm M=m)&\leq&\sum_{j_2=0}^{m-j_1}2^{j_1}{m-j_1\choose j_2}p_e^{j_1}p_s^{j_2}(1-p_s-p_e)^{m-j_1-j_2}\nonumber\\
&=&2^{j_1}p_e^{j_1}(1-p_e)^{m-j_1}.
\end{eqnarray}
Note that in deriving the inequality in~\eqref{eq_Ee}, the summation is taken over the values of $\bfm y$ with $p(\bfm y)\neq 0$. However, in~\eqref{eq_py0} the summation is taken over all possible values of $\bfm y$ of length $m$ (over all $m$-tuples), i.e. $p(\bfm y)=0$ or $p(\bfm y)\neq 0$, which results in the lower bound in~\eqref{eq_py0}. Furthermore, by using the fact that the probability of having $j_1$ erasures in a sequence of length $m$ is equal to ${m\choose j_1 }p_e^{j_1}(1-p_e)^{m-j_1}$, we obtain 
\begin{eqnarray}\label{eq_p(y^{(3)}|y)}
\sum_{\bfm y^{(3)}}p(\bfm y^{(3)}|\bfm M=m)\sum_{\bfm y,p(\bfm y)\neq 0}p(\bfm y^{(3)}|\bfm y,\bfm M=m)&\leq& \sum_{\bfm y^{(3)}}P(d(\bfm y^{(3)})_{e}=j_1|\bfm M=m) 2^{j_1}p_e^{j_1}(1-p_e)^{m-j_1}\nonumber\\
&=&\sum_{j_1=0}^m {m\choose j_1 }p_e^{j_1}(1-p_e)^{m-j_1}(2p_e)^{j_1}(1-p_e)^{m-j_1}\nonumber\\
&=& \left((1-p_e)^2+2p_e^2\right)^m.
\end{eqnarray}
By substituting this result into~\eqref{eq_Hy^{(q)}HY}, we arrive at
\begin{eqnarray}
H(\bfm Y^{(3)})-H(\bfm Y)&\geq&-E\{\bfm M\}\log\left((1+p_e)^2+2p_e^2\right),
\end{eqnarray}
concluding the proof.
\end{IEEEproof}

It is also worth noting that any capacity achieving input distribution over a discrete memoryless channel results in strictly positive output probabilities for possible output sequences of the channel (\cite[p.~95]{gallager_book}). Therefore, for special synchronization error channel models in which for any possible length of the output sequence $m$, all the $m$-tuple output sequences are probable, e.g. i.i.d. deletion channel or i.i.d. random insertion channel, capacity achieving input distributions ($p(\bfm x)$) would result in strictly positive output probability distributions for all $m$-tuple output sequences, i.e. $p(\bfm y^q)>0$ for all $\bfm y^q$ of length $m$ and all possible $m$. Hence, the bounds in~\eqref{eq_py0} and \eqref{eq_p(y^{(3)}|y)} can be thought as equalities for these cases.

\begin{lem}\label{lem_Hy^{(3)}Y|Xe}
In any sub/ers/synch channel and for any input distribution, we have
\begin{equation}\label{eq_H(y^{(3)}|X)}
H({\bfm Y^{(3)}}|\bfm X)\leq H(\bfm Y|\bfm X)+E\left\{\bfm M\right\}H(p_e,p_s,1-p_e-p_s).
\end{equation}
\end{lem}
\begin{IEEEproof}
Due to the result of Lemma~\ref{lem_Hy^{(q)}Y|X} and the fact that in a substitution/erasure channel, regardless of the distribution of $\bfm Y_j$, we can write
\begin{equation}\label{eq_H(y^{(3)}_j|y_j)}
H(Y^{(3)}_j|Y_j)=H(p_e,p_s,1-p_e-p_s),
\end{equation}
hence the proof follows.
\end{IEEEproof}

We can now complete the proof of the main theorem.

\textbf{\textit{Proof of Theorem \ref{thm-synch-sub-ers}}} :
By substituting the results of Lemmas \ref{lem_Hy^{(3)}HYe} and \ref{lem_Hy^{(3)}Y|Xe} into the definition of mutual information, for the same input distribution given to both synchronization error and sub/ers/synch channels, we obtain
\begin{equation}
I(\bfm X;\bfm Y^{(3)})\geq I(\bfm X;\bfm Y)-E\{\bfm M\}\left[H(p_s,p_e,1-p_s-p_e)+\log\left((1+p_e)^2+2p_e^2\right)\right].
\end{equation}
By using the result of Proposition~\ref{prop_I-C}, the proof is completed.\hfill$\blacksquare$

\subsection{Binary Input Symmetric Quaternary Output Channels with Synchronization Errors}
In this subsection, we consider a binary input symmetric quaternary output channel with synchronization errors as described in Section~\ref{sec-channel_model}.
\begin{thm}\label{thm-synch-AWGN4}
The capacity of the binary input symmetric quaternary output channel with synchronization errors $C_{sq}$ can be lower bounded by
\begin{equation}\label{eq_LB-AWGN4}
C_{sq}\geq C_{s}-r\left[H(p_1,p_2,p_3,p_4)+\log\left((p_1+p_3)^2+(p_2+p_4)^2\right)\right],
\end{equation}
where $C_{s}$ denotes the capacity of the synchronization error only channel, and $r$ is as defined in~\eqref{LB_synch-sub-ers}.
\end{thm}

Note that, the presented lower bound is true for all memoryless synchronization error channel models. Therefore, similar to the sub/ers/synch channel we can specialize the results to the Gallager insertion/deletion channel as given in the following corollary.
\begin{corol}
The capacity of binary input symmetric quaternary output channel with insertion/deletion errors (following Gallager's model) $C_{qid}$ is lower bounded by
\begin{equation}\label{eq_LB-ins-del-AWGN4}
C_{qid}\geq C_{id}-(1-p_d+p_i)\left[H(p_1,p_2,p_3,p_4)+\log\left((p_1+p_3)^2+(p_2+p_4)^2\right)\right].
\end{equation} 
\end{corol}

To prove Theorem~\ref{thm-synch-AWGN4}, we need the two lemmas below where the first one gives a lower bound on the output entropy of the binary input quaternary output channel with synchronization errors related to the output entropy of the synchronization error channel, and the second one gives an upper bound on the conditional output entropy of the binary input quaternary output channel with synchronization errors, related to the conditional output entropy of the synchronization error channel.

\begin{lem}\label{lem_HYtildeHYbar}
In any binary input quaternary output channel with synchronization errors and for any input distribution, we have
\begin{equation}
H(\bfm Y^{(4)})\geq H(\bfm Y)-E\left\{\bfm M\right\}\log\left((p_1+p_3)^2+(p_2+p_4)^2\right),
\end{equation}
where $\bfm Y$ denotes the output sequence of the synchronization error channel and input sequence of the binary input quaternary output channel, and $\bfm Y^{(4)}$ denotes the output sequence of the binary input quaternary output channel corresponding to the input sequence $\bfm Y$.
\end{lem}
\begin{IEEEproof}
Similar to the proof of Lemma~\ref{lem_Hy^{(3)}HYe}, we use the result of Lemma~\ref{lem_Hy^{(q)}HY} by taking the summation over all possible sequences of length $m$, i.e., regardless of $p(\bfm y)=0$ or $p(\bfm y)\neq 0$, which results into a looser lower bound. On the other hand, for $p(\bfm y^{(4)}|\bfm y,\bfm M=m)$, we have
\begin{eqnarray}
p(\bfm y^{(4)}|\bfm y,\bfm M=m)&=&\prod_{i=1}^{m}p(Y^{(4)}_i|Y_i)\nonumber\\
&=&p_1^{j_1}p_2^{j_2}p_3^{j_3}p_4^{m-j_1-j_2-j_3},
\end{eqnarray}
where $j_1$ denotes the number of transitions $0\to0^-$ or $1\to1^-$, $j_2$ denotes the number of transitions $0\to0^+$ or $1\to1^+$, and $j_3$ denotes the number of transitions $0\to1^-$ or $1\to0^-$. E.g., $p(0^- 1^+ 0^+ 1^-|0000)=p(0^-|0) p(1^+|0)p( 0^+|0)p( 1^-|0)=p_1p_2p_3p_4$. Furthermore, for a fixed output sequence $\bfm y^{(4)}$ of length $m$ with $j$ $0^-$ symbols, $k$ $0^+$ symbols, $l$ $1^-$ symbols and $m-j-k-l$ $1^+$ symbols, there are ${j\choose i_1}{k\choose i_2}{l\choose i_3}{m-j-k-l\choose i_4}$ possibilities among all $m$-tuples (for $\bfm y$) such that $d(\bfm y,\bfm y^{(4)})_{0\to0^-}=i_1$, $d(\bfm y,\bfm y^{(4)})_{0\to0^+}=i_2$, \mbox{$d(\bfm y,\bfm y^{(4)})_{0\to1^-}=i_3$} and $d(\bfm y,\bfm y^{(4)})_{0\to1^+}=i_4$. By defining $m^-(\bfm y^{(4)})=\#\{t\leq m|y^{(4)}_t\in\{0^-,1^-\}\}$, i.e., the number of the times $y^{(4)}_t=0^-$ or $y^{(4)}_t=1^-$, and $m^+(\bfm y^{(4)})=\#\{t\leq m|y^{(4)}_t\in\{0^+,1^+\}\}$, i.e., the number of the times $y^{(4)}_t=0^+$ or $y^{(4)}_t=1^+$, we can write
\begin{align}
\sum_{\bfm y,p(\bfm y)\neq 0}p(&\bfm y^{(4)}|\bfm y,\bfm M=m)\nonumber\\
&\leq\sum_{i_1=0}^{j}{j\choose i_1}p_1^{i_1}p_3^{j-i_1}\sum_{i_2=0}^{k}{k\choose i_2}p_2^{i_2}p_4^{k-i_2}\sum_{i_3=0}^{l}{l\choose i_3}p_3^{i_3}p_1^{l-i_3}\sum_{i_4=0}^{m-j-k-l}{m-j-k-l\choose i_4}p_4^{i_4}p_2^{m-j-k-l-i_4}\nonumber\\
&=(p_1+p_3)^{j+l}(p_2+p_4)^{m-j-l}\nonumber\\
&=(p_1+p_3)^{m^-(\bfm y^{(4)})}(p_2+p_4)^{m^+(\bfm y^{(4)})}.
\end{align}
By taking the summation over all possible output sequences of length $m$, and using the fact that the probability of having the output $\bfm y^{(4)}$ with length $m$ containing $m^-$ $0^-$ or $1^-$ is ${m\choose m^-}(p_1+p_3)^{m^-}(p_2+p_4)^{m-m^-}$, we obtain
\begin{align}\label{eq_p(y^{(4)}|y)}
\sum_{\bfm y^{(4)}}p(\bfm y^{(4)}|\bfm M=m)\sum_{\bfm y}p(\bfm y^{(4)}|\bfm y,\bfm M=m)&= \sum_{\bfm y^{(4)}}p(\bfm y^{(4)}|\bfm M=m)(p_1+p_3)^{m^-(\bfm y^{(4)})}(p_2+p_4)^{m^+(\bfm y^{(4)})}\nonumber\\
&= \sum_{m^-=0}^m {m\choose m^-}(p_1+p_3)^{m^-}(p_2+p_4)^{m-m^-}(p_1+p_3)^{m^-}(p_2+p_4)^{m-m^-}\nonumber\\
&= \left((p_1+p_3)^2+(p_2+p_4)^2\right)^m
\end{align}
By substituting the result of~\eqref{eq_p(y^{(4)}|y)} into the result of Lemma~\ref{lem_Hy^{(q)}HY}, we obtain
\begin{equation}
H(\bfm Y^{(4)})\geq H(\bfm Y)-E_{\bfm M}\left\{\bfm M\right\}\log\left((p_1+p_3)^2+(p_2+p_4)^2\right),
\end{equation}
which concludes the proof.
\end{IEEEproof}

\begin{lem}\label{lem_Hy^{(4)}Y|X}
For a binary input quaternary output channel with synchronization errors, for any input distribution, we have
\begin{equation}\label{eq_H(y^{(4)}|X)}
H({\bfm Y^{(4)}}|\bfm X)\leq H(\bfm Y|\bfm X)+E_{\bfm M}\left\{\bfm M\right\}H(p_1,p_2,p_3,p_4).
\end{equation}
\end{lem}
\begin{IEEEproof}
Substituting the straightforward result $H(Y^{(4)}_j|Y_j)=H(p_1,p_2,p_3,p_4)$ in the result of Lemma~\ref{lem_Hy^{(q)}Y|X} concludes the proof.
\end{IEEEproof}

We can now complete the proof of Theorem~\ref{thm-synch-AWGN4}.

\textbf{\textit{Proof of Theorem \ref{thm-synch-AWGN4}}} :
Using the results of Lemmas~\ref{lem_HYtildeHYbar} and \ref{lem_Hy^{(4)}Y|X}, we obtain 
\begin{equation}
I(\bfm X;\bfm Y^{(4)})\geq I(\bfm X;\bfm Y)-n r\left [H(p_1,p_2,p_3,p_4)+\log\left((p_1+p_3)^2+(p_2+p_4)^2\right)\right].
\end{equation}
Hence, due the result in Proposition~\ref{prop_I-C}, the proof is complete.\hfill$\blacksquare$

\subsection{Binary Input Symmetric $q$-ary Output Channel with Synchronization Errors (Odd $q$ Case)}
In this subsection, we consider a binary input symmetric $q$-ary output channel with synchronization errors for an arbitrary odd value of $q$, where we represent the transition probability values $P\left(Y^{(q)}_j=k| \bar{Y_j}=b\right)$ for different values of $b\in\{-1,1\}$ and $k=\{-\frac{q-1}{2},\cdots,-1,0,1,\cdots,\frac{q-1}{2}\}$ by $P\left(Y^{(q)}_j=k| \bar{Y}_j=b\right)=p_{k\times b}$. For instance, Table \ref{tb:P(y,y^{(q)})25} shows transition probabilities for a binary input 5-ary output channel.
  \begin{table}[t]
  \centering
  \caption{TRANSITION PROBABILITIES FOR A binary INPUT 5-ARY OUTPUT CHANNEL.}
  \label{tb:P(y,y^{(q)})25}
  \begin{tabular}{|c|c|c|c|c|c|}
  \hline
  \multicolumn{1}{|c|}{} &
  \multicolumn{5}{|c|}{$P(Y^{(q)}_j | \bar{Y}_j)$}\\
  \hline
  $Y_j$ & $Y^{(q)}_j=-2$ & $Y^{(q)}_j=-1$ & $Y^{(q)}_j=0$ & $Y^{(q)}_j=1$&$Y^{(q)}_j=2$\\
  \hline
  $-1$ & $p_2$&$p_1$&$p_0$&$p_{-1}$&$p_{-2}$\\
  \hline
   1 & $p_{-2}$&$p_{-1}$&$p_0$&$p_1$&$p_2$\\
\hline
  \end{tabular}
  \end{table}

The main result on the BSQC channel with synchronization errors with odd $q$ is a generalized version of the result in Theorem~\ref{thm-synch-sub-ers}.
\begin{thm}\label{thm-synch-KQodd}
The capacity of the BSQC channel with synchronization errors $C_{Qs}$ for an odd $q$ can be lower bounded by
\begin{equation}\label{eq_LB-KQodd}
C_{Qs}\geq C_{s}-r\left(H(p_{-\frac{q-1}{2}},\cdots ,p_{\frac{q-1}{2}})+\log\left(2p_0^2+\sum_{k=1}^{\frac{q-1}{2}}(p_k+p_{-k})^2\right)\right),
\end{equation}
where $C_{s}$ denotes the capacity of the binary input synchronization error channel.
\end{thm}
\begin{IEEEproof}
The proof of the theorem is given in Appendix~\ref{app_odd}.
\end{IEEEproof}

\subsection{Binary Input Symmetric $q$-ary Output Channel with Synchronization Errors (Even $q$ Case)}

We now consider the generalization of the result of Theorem~\ref{thm-synch-AWGN4} for even $q$. For the transition probabilities of the binary input $q$-ary output channel, we define $P\left(Y^{(q)}_j=k| \bar{Y_j}=b\right)=p_{k\times b}$, where $b\in\{-1,1\}$ and $k=\{-\frac{q}{2},\cdots,-1,1,\cdots,\frac{q}{2}\}$. For instance, Table \ref{tb:P(y,y^{(q)})26} shows transition probabilities for a binary input 6-ary output channel.
  \begin{table}[t]
  \centering
  \caption{Transition probabilities for a binary input symmetric 6-ary output channel.}
  \label{tb:P(y,y^{(q)})26}
  \begin{tabular}{|c|c|c|c|c|c|c|}
  \hline
  \multicolumn{1}{|c|}{} &
  \multicolumn{6}{|c|}{$P(Y^{(q)}_j | \bar{Y}_j)$}\\
  \hline
  $Y_j$ & $Y^{(q)}_j=-3$ & $Y^{(q)}_j=-2$ & $Y^{(q)}_j=-1$ & $Y^{(q)}_j=1$&$Y^{(q)}_j=2$&$Y^{(q)}_j=3$\\
  \hline
  $-1$ & $p_3$&$p_2$&$p_1$&$p_{-1}$&$p_{-2}$&$p_{-3}$\\
  \hline
   1 & $p_{-3}$&$p_{-2}$&$p_{-1}$&$p_1$&$p_2$&$p_3$\\
\hline
  \end{tabular}
  \end{table}

The main result on the BSQC channel with synchronization errors for any $q$ is given in the following theorem.
\begin{thm}\label{thm-synch-KQeven}
Capacity of the BSQC channel with synchronization errors $C_{Qs}$, for any even $q$ can be lower bounded by
\begin{equation}\label{eq_LB-KQeven}
C_{Qs}\geq C_{s}-r\left[H(p_{-\frac{q}{2}},\cdots ,p_{-1},p_1,\cdots ,p_{\frac{q}{2}})+\log\left(\sum_{k=1}^{\frac{q}{2}}\left(p_{k}+p_{-k}\right)^2\right)\right],
\end{equation}
where $C_{s}$ denotes the capacity of the binary input synchronization error channel.
\end{thm}
\begin{IEEEproof}
The proof of Theorem \ref{thm-synch-KQeven} is given in Appendix~\ref{app_even}.
\end{IEEEproof}

\section{Achievable Rates over BI-AWGN Channels with Synchronization Errors}\label{sec-AWGN}
In this section, a binary synchronization error channel in the presence of AWGN is considered as defined in Section~\ref{sec-AWGN_model}. We present two different lower bounds on the capacity of the AWGN/synch channel.

Before giving the main results on AWGN/synch channel, we would like to make some comments on its information stability.

\subsection{Information Stability of Memoryless Discrete Input Continuous Output Channels with Synchronization Errors}
It is shown in~\cite{dobrushin_general} that the Shannon's theorem holds in any information stable channel. In~\cite{dobrushin}, the information stability of the memoryless discrete input discrete output channels with synchronization errors is proved which shows that the Shannon's theorem holds in such a channel. It can be observed that the proofs used in~\cite{dobrushin} can be also generalized to the continuous output case as discussed in this section.

To prove the information stability, it is sufficient to prove the existence of the limit
\begin{equation}
C=\lim_{N\to\infty}\frac{1}{N}C_{N}=\lim_{N\to\infty}\frac{1}{N}\max_{P(\bfm X)}I(\bfm X;\widetilde{\bfm Y}),
\end{equation}
which is the information capacity of the channel, and the existence of an information stable sequence of two random variables $\left(\bfm X,\widetilde{\bfm Y}\right)$, which achieves the capacity of the channel. 

The only difference between the channel considered here with the channel considered by Dobrushin in~\cite{dobrushin}, is that in the continuous output case the output symbols belong to an infinite set. However, this difference does not have any effect on the steps of proofs. The existence of the limit in \cite[Section~IV]{dobrushin} is proved based on the memoryless property of the channel which also holds in the continuous output case. 

In the case of the existence of an information stable sequence achieving the capacity (\cite[Section~V]{dobrushin}), there is no need to condition on the discrete output symbol values, and all the reasoning hold for the continuous output case as well.  
The key point in the proof is that the channel is stationary which also holds for the continuous output case, such that the same genie-aided channel as the one considered for the discrete output channel can also be considered for our case. The genie-aided channel is obtained by inserting markers through the transmission after transmitting each block of length $k$, where the entire length of transmission is $K=gk+l$ ($l<k$). 

The other point in the proof is the number of possibilities in converting the output of the original channel $\widetilde{\bfm Y}$ into the output of the genie-aided channel $\widetilde{\bfm Y}'$, i.e., $|f^{-1}(\widetilde{Y})|$ where $\widetilde{Y}=f(\widetilde{Y}')$. Since for the continuous output case we still have $\displaystyle\lim_{g\to\infty}\frac{\max_{\widetilde{Y}} |f^{-1}(\widetilde{Y})|}{g}\to 0$, the proof holds. 

Since, both capacity convergence and existence of an information stable sequence which achieves the capacity remain valid in the continuous output case as well, we can conclude that the memoryless discrete input continuous output channels with synchronization errors are also information stable and, as a result, the Shannon's theorem applies in such a channel.

\subsection{Capacity Lower Bounds for AWGN/Synch Channels}
Here, we present two results on the capacity of an AWGN/synch channel. Both results are generalizations of results for the discrete output channels when the number of quantization levels goes to infinity. The first result is obtained by employing a uniform quantizer while in deriving the second result a non-uniform quantizer is employed which provides a tighter lower bound.

\begin{thm}\label{thm-AWGN-quan}
Capacity of the AWGN/synch channel $C_{As}$ can be lower bounded by
\begin{equation}\label{eq_LB-uniform}
C_{As}\geq C_{s}-r\log\left(\sqrt{\frac{e}{2}}(1+e^{-\frac{1}{\sigma^2})}\right),
\end{equation}
where $C_{s}$ denotes the capacity of the synchronization error channel.
\end{thm}

We give an outline of the proof and defer its details to Appendix~\ref{app_quan1}. We consider a quantized version of the output symbols via a $2M$-levels uniform quantizer by quantization intervals of length $\Delta$ with $M$ going to infinity and $\Delta$ going to zero. Therefore, for $p_m$ ($m=\{-M,\cdots,-1,1,\cdots,M\}$) which denotes the probability that the continuous output symbol, $\widetilde{Y}_j$, being quantized to the $bm$-th quantization level ($b\in\{-1,1\}$) conditioned on $\bar{X}_j=b$ being transmitted, we obtain
\begin{equation}\label{eq_pi}
p_m= \left\{\begin{array}{ccc}
Q(\frac{1-m\Delta}{\sigma})-Q(\frac{1-(m-1)\Delta}{\sigma})&,& m > 0\\
Q(\frac{1+(|m|-1)\Delta}{\sigma})-Q(\frac{1+|m|\Delta}{\sigma})&,& m < 0
\end{array}\right.,
\end{equation}
where $Q(.)$ is the right tail probability of the standard normal distribution. By substituting~\eqref{eq_pi} into the result of Theorem~\ref{thm-synch-KQeven}, we can write 
\begin{equation}
C_{As}\geq C_{s}-r\lim_{M\to\infty, \Delta\to 0}\left[H(p_{-M},\cdots,p_{-1},p_1,\cdots, p_{M})+\log\left(\sum_{m=1}^{M}\left(p_{m}+p_{-m}\right)^2\right)\right].
\end{equation}
Finally, by using the fact that when $M\to\infty$ and $\Delta\to 0$, we have $p_m\cong f(1-m\Delta)\Delta$, with $f(x)=\frac{1}{\sqrt{2\pi}\sigma}e^{-\frac{x^2}{2\sigma^2}}$, after some algebra (detailed in Appendix~\ref{app_quan1}), we obtain the result in~\eqref{eq_LB-uniform}.


This result is obtained as a straightforward generalization of the discrete output channel results by employing a symmetric uniform quantizer, but the result may not be tight. For instance, for $\sigma=0$, i.e. the noiseless scenario, the result does not match with the trivial result which is $C_{As}=C_s$ for $\sigma=0$. We expect that if we apply an appropriate non-uniform quantizer on the output symbols of the AWGN/synch channel, we can achieve a tighter lower bound on its capacity (which also agrees with the trivial result for $C_{As}=C_s$ for $\sigma=0$). By using this idea, we present our main result on the capacity of an AWGN/synch channel in the following theorem by using a symmetric non-uniform quantizer.

\begin{thm}\label{thm-AWGN-quan-nonunifrom}
Let $C_s$ denote the capacity of the synchronization error channel, then for the capacity of the AWGN/synch channel $C_{As}$, we obtain
\begin{equation}\label{eq_LB-quan}
C_{As}\geq C_s-r\left[\log(e)\left(\frac{2}{\sqrt{2\pi}\sigma}e^{-\frac{1}{2\sigma^2}}-\frac{2}{\sigma^2}Q\left(\frac{1}{\sigma}\right)\right)+\log\left(1+Q\left(\frac{1}{\sigma}\right)+e^{\frac{4}{\sigma^2}}Q\left(\frac{3}{\sigma}\right)\right)\right].
\end{equation}
\end{thm}

\begin{IEEEproof}
To prove the theorem, we first define an appropriate symmetric non-uniform quantizer with $2M$ quantization levels. Then, by letting $M$ go to infinity and employing the result of Theorem~\ref{thm-synch-KQeven}, we complete the proof. 

In general, by utilizing any symmetric quantizer with $2M$ quantization levels on the output symbols $\widetilde{Y}_j$, for the transition probabilities of the resulting binary input symmetric $2M$-ary output channel, we have
\begin{equation}
p_m=P(Y^{(2M)}=bm|\bar{X_j}=b)=\left\{\begin{array}{ccc}
P(t_{m-1}<\widetilde{Y}_j<t_{m})&,& 0<m\leq M\\
P(-t_{m}<\widetilde{Y}_j<-t_{m-1})&,& -M\leq m< 0
\end{array}\right.,
\end{equation}
where $t_{-m}=-t_{m}$, $t_0=0$ and $t_{m-1}<t_m$ for $m=\{1,\cdots,M\}$. We choose the quantization step sizes, i.e., $\Delta_m=t_{m}-t_{m-1}$ for $m=\{1,\cdots,M\}$, to satisfy $p_1=p_2=\cdots=p_M$. Note that due to symmetry of the quantizer $\Delta_{-m}=\Delta_m$ (as illustrated in Fig.~\ref{fig_nonuniform}). 
\begin{figure}
    \begin{center}
    \includegraphics[trim=28mm 105mm 20mm 110mm,clip,  width=.53\textwidth]{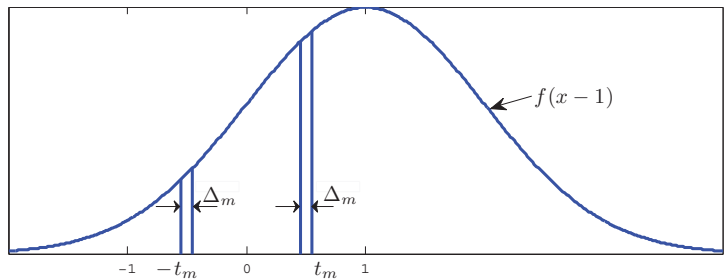}
    \end{center}
    \caption{Symmetric non-uniform quantizer step sizes.}
    \label{fig_nonuniform}
\end{figure}
On the other hand, by defining $P=Q(\frac{1}{\sigma})$, we have $\displaystyle\sum_{m=1}^{M}p_{-m}=P$ and $\displaystyle\sum_{m=1}^{M}{p_m}=1-P$ which results in $p_m=\frac{1-P}{M}$ for $m=\{1,\cdots,M\}$.

Using the result of Theorem~\ref{thm-synch-KQeven}, to derive a lower bound on the capacity of the channel with $2M$-level quantized outputs, we need to obtain 
$H(p_{-M},\cdots,p_{-1},p_{1},\cdots,p_{M})+\log\left(\sum_{m=1}^M(p_m+p_{-m})^2\right)$. In the following, we first compute the exact values of $H_M=H(p_{-M},\cdots,p_{-1},p_{1},\cdots,p_{M})-\log(M)$ and $\log\left(\sum_{m=1}^M(p_m+p_{-m})^2\right)+\log(M)$. For $H_M$, we have
\begin{align}\label{eq_H}
H_M&=-\sum_{m=1}^{M}p_m\log(p_m)-\sum_{m=1}^{M}p_{-m}\log(p_{-m})-\log(M)\nonumber\\
&=-(1-P)\log(1-P)-\sum_{m=1}^{M}p_{-m}\log(Mp_{-m}).
\end{align}
To calculate $-\sum_{m=1}^{M}p_{-m}\log(Mp_{-m})$, we first derive a relation between $p_m$ and $p_{-m}$ by using the fact that $\Delta_m=\Delta_{-m}$. For large $M$ and $m=\{1,\cdots,M\}$, we have $p_m\cong f(1-t_m)\Delta_m$ and $p_{-m}\cong f(1+t_m)\Delta_m$, where $f(x)=\frac{1}{\sqrt{2\pi}\sigma}e^{-\frac{x^2}{2\sigma^2}}$. Furthermore, since $p_m=\frac{1-P}{M}$ for $m=\{1,\cdots,M\}$ and $\frac{f(1+t_m)}{f(1-t_m)}\cong e^{-\frac{2t_m}{\sigma^2}}$, we can write
\begin{align}
p_{-m}&\cong\frac{f(1+t_m)}{f(1-t_m)}p_m\nonumber\\
&=\frac{1-P}{M}e^{-\frac{2t_m}{\sigma^2}},
\end{align}
with the understanding that the approximation becomes exact as $M\to\infty$.
By using this result, we obtain
\begin{align}\label{eq_H1}
\lim_{M\to\infty}-\sum_{m=1}^{M}p_{-m}\log(Mp_{-m})&=\lim_{M\to\infty}-\sum_{m=1}^{M}p_{-m}\log\left(1-P\right)-\lim_{M\to\infty}\sum_{m=1}^{M}p_{-m}\log\left(e^{-\frac{2t_m}{\sigma^2}}\right)\nonumber\\
&=-P\log\left(1-P\right)-\lim_{M\to\infty}\sum_{m=1}^{M}p_{-m}\log\left(e^{-\frac{2t_m}{\sigma^2}}\right),
\end{align} 
where we used the fact that $\sum_{m=1}^Mp_{-m}=P$. On the other hand, for $\displaystyle\lim_{M\to\infty}-\sum_{m=1}^{M}p_{-m}\log\left(e^{-\frac{2t_m}{\sigma^2}}\right)$, we can write
\begin{align}\label{eq_H2}
\lim_{M\to\infty}-\sum_{m=1}^{M}p_{-m}\log\left(e^{-\frac{2t_m}{\sigma^2}}\right)&=\lim_{M\to\infty}\log(e)\sum_{m=1}^{M}f(1+t_m)\Delta_m\frac{2t_m}{\sigma^2}\nonumber\\
&=\log(e)\int_0^\infty f(1+t)\frac{2t}{\sigma^2}dt\nonumber\\
&=\log(e)\frac{2}{\sigma^2}\int_0^\infty \frac{t}{\sqrt{2\pi}\sigma}e^{-\frac{(t+1)^2}{2\sigma^2}}dt\nonumber\\
&=\log(e)\left(\frac{2}{\sqrt{2\pi}\sigma}e^{-\frac{1}{2\sigma^2}}-\frac{2}{\sigma^2}P\right).
\end{align}
By substituting~\eqref{eq_H2} and \eqref{eq_H1} into~\eqref{eq_H}, we obtain
\begin{equation}\label{eq_HH}
\lim_{M\to\infty}H_M=-\log(1-P)+\log(e)\left(\frac{2}{\sqrt{2\pi}\sigma}e^{-\frac{1}{2\sigma^2}}-\frac{2}{\sigma^2}P\right).
\end{equation}
At this point, we only need to obtain the exact value of $\sum_{m=1}^M(p_m+p_{-m})^2$, where we have
\begin{align}
\sum_{m=1}^MM(p_m+p_{-m})^2&=\sum_{m=1}^MM(p_m^2+2p_mp_{-m}+p_{-m}^2)\nonumber\\
&=(1-P)^2+2P(1-P)+\sum_{m=1}^MMp_{-m}^2.
\end{align}
Furthermore, if we let $M$ go to infinity, for $\sum_{m=1}^MMp_{-m}^2$, we can write
\begin{align}\label{eq_logp^2}
\lim_{M\to\infty}\sum_{m=1}^MMp_{-m}^2&=\lim_{M\to\infty}\sum_{m=1}^MMf(1+t_m)\Delta_m \frac{f(1+t_m)}{f(1-t_m)}p_m\nonumber\\
&=\lim_{M\to\infty}(1-P)\sum_{m=1}^M  \frac{1}{\sqrt{2\pi}\sigma}e^{-\frac{(t_m+1)^2}{2\sigma^2}}e^{-\frac{2t_m}{\sigma^2}}\Delta_m\nonumber\\
&=(1-P)\int_0^\infty \frac{1}{\sqrt{2\pi}\sigma}e^{-\frac{(t+1)^2}{2\sigma^2}}e^{-\frac{2t}{\sigma^2}}dt\nonumber\\
&=(1-P)\int_0^\infty \frac{1}{\sqrt{2\pi}\sigma}e^{-\frac{(t+3)^2-8}{2\sigma^2}}dt\nonumber\\
&=(1-P) e^{\frac{4}{\sigma^2}}Q\left(\frac{3}{\sigma}\right).
\end{align}
Using the results of~\eqref{eq_logp^2} and \eqref{eq_HH}, we obtain
\begin{align}
\lim_{M\to\infty}\bigg(H(p_{-M},\cdots,p_{-1},p_{1},\cdots,p_{M})&+\log\left(\sum_{m=1}^M(p_m+p_{-m})^2\right)\bigg)= \lim_{M\to\infty}\left(H_M+\log\left(\sum_{m=1}^MM(p_m+p_{-m})^2\right) \right)\nonumber\\
&=\log(e)\left(\frac{2e^{-\frac{1}{2\sigma^2}}}{\sqrt{2\pi}\sigma}-\frac{2}{\sigma^2}P\right)+\log\left(1+P+e^{\frac{4}{\sigma^2}}Q\left(\frac{3}{\sigma}\right)\right)
\end{align}
Finally, by substituting this result into~\eqref{eq_LB-KQeven}, the proof follows.
\end{IEEEproof}\color{black} 
By employing a symmetric non-uniform quantizer, we achieve a tighter lower bound on the capacity of the AWGN/synch channel compared to the lower bound in Theorem~\ref{thm-AWGN-quan}. The result is also in agreement with the trivial result $C_{As}=C_s$ ($\sigma=0$). A primary advantage of the derived lower bound in~\eqref{eq_LB-quan} is that we can use any lower bound on the capacity of the synchronization error only channel to lower bound the capacity of the AWGN/synch channel.

\section{Numerical Examples}\label{sec-num}
In this section, we give several numerical examples of the lower bounds on the capacity of the ins/del/sub and del/AWGN channel and compare them with the existing ones in the literature. To the best of our knowledge, there are no existing results on lower bounding the capacity of the ins/del/sub/ers and ins/del/AWGN channels, therefore, our results will provide a benchmark for these general cases.

\subsection{Insertion/Deletion/Substitution Channel}
In Table~\ref{tb_ins-del-sub}, we compare the lower bound on the capacity of the ins/del/sub channel~\eqref{LB_ins-del-sub} with the existing lower bounds in~\cite{gallager,dario2} for several values of $p_d$, $p_i$ and $p_s$. We employ the lower bound derived in~\cite{drinea2007} as the lower bound on the capacity of the deletion channel and the lower bound in~\cite{dario2} as the lower bound on the capacity of the ins/del channel in~\eqref{LB_ins-del-sub}. 
Note that the Gallager's model in \cite{gallager} by parameters $p_d$, $p_i$ and $p_c$ can be considered as concatenation of an ins/del channel with parameters $p_d$ and $p_i$, and a BSC channel with cross error probability of $p_s$ where $p_s$ is the solution of $p_c=(1-p_s)(1-p_d-p_i)$. The advantage of the lower bound~\eqref{LB_ins-del-sub} is in using the tightest lower bound on the capacity of the ins/del channel in lower bounding the capacity of the overall channel, i.e., the information rate of the overall channel is lower bounded for the input distribution which results in the tightest lower bound on the capacity of the ins/del channel. 
We observe that for $p_i=0$, a fixed $p_d$ and small values of $p_s$, the lower bound~\eqref{LB_ins-del-sub} improves the lower bound given in~\cite{dario2}. 
This is not unexpected, because for small values of $p_s$ the input distribution achieving the capacity of the i.i.d. deletion channel is not far from the optimal input distribution of the del/sub channel. We also observe that the lower bound~\eqref{LB_ins-del-sub} outperforms the lower bound given in~\cite{gallager}. However, for the case $p_i\neq 0$ it does not improve the lower bound given in~\cite{dario2}, since as the lower bound on the capacity of ins/del channel we used the result in~\cite{dario2} and lower bounded further to achieve lower bound on the capacity of the overall channel. 
\begin{table}[h]
    \centering
    \caption{Comparing the lower bound derived on the capacity of the ins/del/sub channel with existing lower and upper bounds (Boldface numbers show the best bounds).}
    \begin{tabular}{|c|c|c|c|c|c|c|}
        \hline
        $p_d$ & $p_i$ & $p_s$ & LB from~\cite{gallager} & 
        LB~\eqref{LB_ins-del-sub}  & LB from~\cite{dario2}& UB 				 from~\cite{dario2}\\
		\hline\hline
		0.001&0.00&0.001&  0.9772&\bf{0.9775} &0.9773& 0.9856\\
		\hline
		0.001&0.00&0.01&0.9079&\bf{0.9082} &0.9081&0.9163 \\
		\hline
		0.001&0.00&0.1& 0.5201& 0.5204 & \bf{0.5210}& 0.5292   \\ 
		\hline
		0.01& 0.00& 0.001&0.9079&\bf{0.9107}&0.9091&0.9586\\
		\hline
        0.01& 0.00& 0.01& 0.839& \bf{0.842}& \bf{0.842}& 0.886\\
        \hline
        0.01& 0.00& 0.10& 0.454& 0.458& \bf{0.466}& 0.510\\
        \hline
        0.10&0.000&0.001&0.5207&\bf{0.5514}&0.5346&0.7300\\
        \hline
        0.10& 0.00& 0.01& 0.458& 0.489& \bf{0.492}& 0.644\\
        \hline
        0.10& 0.00& 0.10& 0.108& 0.140& \bf{0.211}& 0.363\\
        \hline
        0.10&0.10&0.001&0.0689&0.1678 &\bf{0.1761}&0.4504\\
        \hline
        0.10& 0.10& 0.01& 0.013& 0.0984& \bf{0.139}& 0.438\\
        \hline
    \end{tabular} 		
    \label{tb_ins-del-sub}
\end{table}

\subsection{Insertion/Deletion/AWGN Channel}
We now give several numerical examples of the lower bound~\eqref{eq_LB-quan} on the capacity of the ins/del/AWGN channel and compare them with existing results. In the literature, there are only a few results on the capacity of the deletion/AWGN channel, e.g., the simulation based bound of~\cite{junhu} which is the achievable information rate of the deletion/AWGN channel for i.u.d. input sequences obtained by Monte-Carlo simulations and the analytical result given in~\cite{IT-paper} which is a lower bound on the information rate for i.u.d. input sequences, and no previous results are available for the ins/del/AWGN case.

\begin{figure}
    \centering
    \includegraphics[trim=10mm 65mm 10mm 60mm,clip,  width=.7\textwidth]{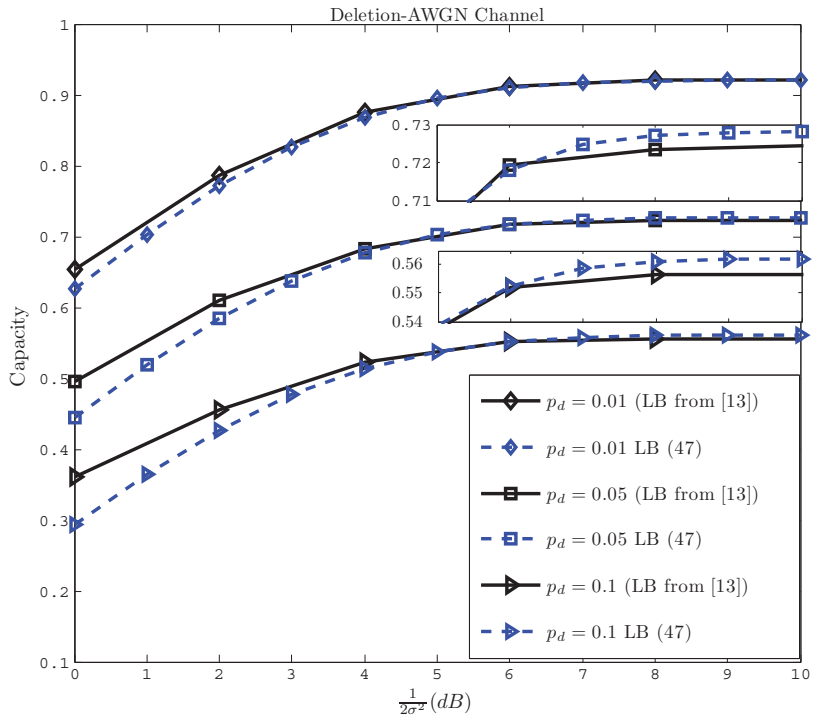}
    \caption{Comparison between the lower bound \eqref{eq_LB-quan} with the lower bound in~\cite{junhu} versus SNR for different deletion probabilities.}
    \label{fig:AWGN}
\end{figure}

Fig.~\ref{fig:AWGN} shows a comparison of the lower bound on the capacity of the del/AWGN channel in~\eqref{eq_LB-quan} with the results in~\cite{junhu}. We observe from Fig.~\ref{fig:AWGN} that the lower bound~\eqref{eq_LB-quan} is far away from the simulation based results of~\cite{junhu} for large $\sigma^2$ values and small deletion probabilities. This is not unexpected, because in~\cite{junhu}, the achievable information rate for i.u.d. input sequences are obtained (through lengthy Monte-Carlo simulations) and i.u.d. inputs are close to optimal. However, the procedure employed in~\cite{junhu} is only useful for computing capacity lower bounds for small values of deletion probabilities, e.g. $p_d\leq0.1$, while the lower bound in~\eqref{eq_LB-AWGN4} holds for the entire range of deletion probabilities by employing any lower bound on the capacity of the deletion channel in lower bounding the capacity of the deletion/AWGN channel. We also observe that, since in deriving the lower bound~\eqref{eq_LB-quan} on the capacity of the deletion/AWGN channel, we employ the tightest lower bound presented on the capacity of the deletion channel, for small values of $\sigma^2$, the lower bound~\eqref{eq_LB-quan} improves the lower bound given in~\cite{junhu}.

\section{Summary And Conclusions}\label{sec-conc}
In this paper, we presented several lower bounds on the capacity of binary input symmetric output channels with synchronization errors in addition to substitutions, erasures or AWGN. We showed that the capacity of any channel with synchronization errors which can be considered as a cascade of two channels (where only the first one suffers from synchronization errors and the second one is a memoryless channel) can be lower bounded in terms of the capacity of the first channel and the parameters of the second channel. We considered two classes of channels: binary input symmetric $q$-ary output channels (e.g., for $q=3$ a binary input channel with substitutions and erasures) with synchronization errors and BI-AWGN channels with synchronization errors. We gave the first lower bound on the capacity of substitution/erasure channel with synchronization errors and the first analytical result on the capacity of BI-AWGN channel with synchronization errors. We also demonstrated that the lower bounds developed on the capacity of the del/AWGN channel for small $\sigma^2$ values and the del/sub channel for small values of $p_s$ improve the existing results.

\appendices
\section{Proof of Theorem \ref{thm-synch-KQodd}}\label{app_odd}
We first give a lower bound on the output entropy of the binary input $q$-ary output channel with synchronization errors related to the output entropy of the binary synchronization error channel, then give an upper bound on the conditional output entropy of the binary input $q$-ary output channel with synchronization errors related to the conditional output entropy of the binary synchronization error channel.

\begin{lem}\label{lem_HYkqodd}
For a binary input $q$-ary output channel with synchronization errors, for any input distribution and any odd $q$, we have
\begin{equation}
H(\bfm Y^{(q)})\geq H(\bfm Y)-E\{\bfm M\}\log\left(2p_0^2+\sum_{k=1}^{\frac{q-1}{2}}(p_k+p_{-k})^2\right),
\end{equation}
where $\bfm Y$ denotes the output sequence of the synchronization error channel and input sequence of the binary input symmetric $q$-ary output channel, and $\bfm Y^{(q)}$ denotes the output sequence of the binary input symmetric $q$-ary output channel.
\end{lem}

\begin{IEEEproof}
For $p(\bfm y^{(q)}|\bfm y,\bfm M=m)$, we have
\begin{equation}
p(\bfm y^{(q)}|\bfm y,\bfm M=m)=\prod_{k=-\frac{q-1}{2}}^{\frac{q-1}{2}} p_k^{j_k},
\end{equation}
where $j_k$ denotes the number of transitions $b\to \frac{k}{b}$. E.g., in a binary input 5-ary output channel we have $p(-1102|1111)=p_{-1}p_1p_0p_2$. Therefore, for a fixed output sequence $\bfm y^{(q)}$ of length $m$ with $j_{k}$ symbols of $k$, since there are $2^{j_0}\prod_{k=1}^{\frac{q-1}{2}}{j_k\choose i_k}{j_{-k}\choose i_{-k}}$ possibilities for $\bfm y$ such that $d(\bfm y,\bfm y^{(q)})_{b\to 0}=j_0$ and $d(\bfm y,\bfm y^{(q)})_{b\to \frac{k}{b}}=i_k$, we can write
\begin{eqnarray}
\sum_{\bfm y,p(\bfm y\neq 0)}p(\bfm y^{(q)}|\bfm y,\bfm M=m)&\leq &2^{j_0}p_0^{j_0}\prod_{q=1}^{\frac{q-1}{2}}\sum_{i_k=0}^{j_k}{j_k\choose i_k}p_k^{i_k}p_{-k}^{j_k-i_k}\sum_{i_{-k}=0}^{j_{-k}}{j_{-k}\choose i_{-k}}p_{-k}^{i_{-k}}p_{k}^{j_{-k}-i_{-k}}\nonumber\\
&=&2^{j_0}p_0^{j_0}\prod_{k=1}^{\frac{q-1}{2}}(p_k+p_{-k})^{j_k+j_{-k}}\nonumber\\
&=&2^{m_0}p_0^{m_0}\prod_{k=1}^{\frac{q-1}{2}}(p_k+p_{-k})^{m_k(\bfm y^{(q)})},
\end{eqnarray}
where $m_k(\bfm y^{(q)})=\#\{t\leq m|y^{(q)}_t\in\{k,-k\}\}$, i.e., the number of the times $Y^{(q)}_t=k$ or $Y^{(q)}_t=-k$. Hence,
\begin{align}\label{eq_p(y^{(q)}|y)odd}
\sum_{\bfm y^{(q)}}p(\bfm y^{(q)}|\bfm M=m)\sum_{\bfm y,p(\bfm)\neq 0}&p(\bfm y^{(q)}|\bfm y,\bfm M=m)\leq \sum_{\bfm y^{(q)}}p(\bfm y^{(q)}|\bfm M=m)(2p_0)^{m_0}\prod_{k=1}^{\frac{q-1}{2}}(p_k+p_{-k})^{m_k(\bfm{y}^{(q)})}\nonumber\\
&= \sum_{m_0+\cdots +m_{\frac{q-1}{2}}=m} {m\choose m_0,\cdots, m_\frac{q-1}{2}}p_0^{m_0}\prod_{l=1}^{\frac{q-1}{2}}(p_l+p_{-l})^{m_l}\left((2p_0)^{m_0}\prod_{k=1}^{\frac{q-1}{2}}(p_k+p_{-k})^{m_k}\right)\nonumber\\
&= \left(2p_0^2+\sum_{k=1}^{\frac{q-1}{2}}(p_k+p_{-k})^2\right)^m.
\end{align}
By substituting the result of~\eqref{eq_p(y^{(q)}|y)odd} in the result of Lemma~\ref{lem_Hy^{(q)}HY}, we obtain
\begin{eqnarray}
H(\bfm Y^{(q)})&\geq& H(\bfm Y)-E\{\bfm M\}\log\left(2p_0^2+\sum_{k=1}^{\frac{q-1}{2}}(p_k+p_{-k})^2\right)\nonumber\\
&=&-E\{\bfm M\}\log\left(2p_0^2+\sum_{k=1}^{\frac{q-1}{2}}(p_k+p_{-k})^2\right),
\end{eqnarray}
which concludes the proof.
\end{IEEEproof}

\begin{lem}\label{lem_HY^{(q)}Y|Xodd}
For a binary input $q$-ary output channel with synchronization errors, for any odd $q$ and any input distribution, we have
\begin{equation}\label{eq_H(Y'|X)}
H({\bfm Y^{(q)}}|\bfm X)\leq H(\bfm Y|\bfm X)+E\{\bfm M\}H(p_{-\frac{q-1}{2}},\cdots ,p_{\frac{q-1}{2}}).
\end{equation}
\end{lem}
\begin{IEEEproof}
By using the result of Lemma~\ref{lem_Hy^{(q)}Y|X}, we can write
\begin{eqnarray}
H(\bfm Y^{(q)}|\bfm X)&\leq& E\{\bfm M\} H(Y^{(q)}_j|Y_j)+H(\bfm Y|\bfm X)\nonumber\\
&=&E\{\bfm M\}H(p_{-\frac{q-1}{2}},\cdots ,p_{\frac{q-1}{2}})+H(\bfm Y|\bfm X).
\end{eqnarray}
\end{IEEEproof}

Obviously, by employing the results of Lemmas~\ref{lem_HYkqodd} and \ref{lem_HY^{(q)}Y|Xodd} and using the same approach as in the proof of Theorem~\ref{thm-synch-sub-ers}, the proof of Theorem~\ref{thm-synch-KQodd} is complete. 
\section{Proof of Theorem \ref{thm-synch-KQeven}}\label{app_even}
We need the following two lemmas to proof Theorem~\ref{thm-synch-KQeven}. In the first one, a lower bound on the output entropy of the binary input $q$-ary output channel with synchronization errors is derived relating with the output entropy of the binary synchronization error channel. In the second one, we give an upper bound on the conditional output entropy of the binary input $q$-ary output channel with synchronization errors related to the conditional output entropy of the binary synchronization error channel. By employing the result of two following lemmas and using the same approach as in the proof of Theorem~\ref{thm-synch-AWGN4}, Theorem~\ref{thm-synch-KQeven} is proved.

\begin{lem}\label{lem_HYkqeven}
For a binary input $q$-ary output channel with synchronization errors, for any input distribution and any even $q$, we have
\begin{equation}
H(\bfm Y^{(q)})\geq H(\bfm Y)-E\{m\}\log\left(\sum_{k=1}^{\frac{q}{2}}\left(p_{k}+p_{-k}\right)^2\right),
\end{equation}
where $\bfm Y$ denotes the output sequence of the synchronization error channel and input sequence of the binary input symmetric $q$-ary output channel, and $\bfm Y^{(q)}$ denotes the output sequence of the binary input $q$-ary output channel.
\end{lem}

\begin{IEEEproof}
Due to the result of Lemma~\ref{lem_Hy^{(q)}HY}, we have
\begin{equation}\label{eq_Hy^{(q)}-H-2Q}
H(\bfm Y^{(q)})-H(\bfm Y)\geq -E_{\bfm M}\left\{\log\left(\sum_{\bfm y^{(q)}}\sum_{\bfm y,p(\bfm y\neq 0)}p(\bfm y^{(q)}|\bfm y,\bfm M=m)p(\bfm y^{(q)}|\bfm M=m)\right)\right\}.
\end{equation}
On the other hand for $p(\bfm y^{(q)}|\bfm y,\bfm M=m)$, we have
\begin{equation}
p(\bfm y^{(q)}|\bfm y,\bfm M=m)=\prod_{1}^{\frac{q}{2}} p_k^{j_k} p_{-k}^{j_{-k}},
\end{equation}
where $j_k$ denotes the number of transitions $b\to \frac{k}{b}$. For instance, in a binary input 6-ary output channel we have $p(-11-32|1111)=p_{-1}p_1p_{-3}p_2$. On the other hand, for a fixed output sequence $\bfm y^{(q)}$ of length $m$ with $j_{k}$ symbols of $k$, there are $\prod_{k=1}^{\frac{q}{2}}{j_k\choose i_k}{j_{-k}\choose i_{-k}}$ possibilities for $\bfm y$ such that $d(\bfm y,\bfm y^{(q)})_{b\to \frac{k}{b}}=i_k$. By defining \mbox{$m_k(\bfm y^{(q)})=\#\{t\leq m|y^{(q)}_t\in\{k,-k\}\}$}, i.e., the number of the times $Y^{(q)}_t=k$ or $Y^{(q)}_t=-k$, we can write
\begin{eqnarray}\label{eq_above}
\sum_{\bfm y,p(\bfm y\neq 0)}p(\bfm y^{(q)}|\bfm y,\bfm M=m)&\leq&\prod_{k=1}^{\frac{q}{2}}\sum_{i_k=0}^{j_k}{j_k\choose i_k}p_k^{i_k}p_{-k}^{j_k-i_k}\sum_{i_{-k}=0}^{j_{-k}}{j_{-k}\choose i_{-k}}p_{-k}^{i_{-k}}p_{-k}^{j_{-k}-i_{-k}}\nonumber\\
&=&\prod_{k=1}^{\frac{q}{2}}(p_k+p_{-k})^{j_k+j_{-k}}\nonumber\\
&=&\prod_{k=1}^{\frac{q}{2}}(p_k+p_{-k})^{m_k(\bfm y^{(q)})},
\end{eqnarray}
Furthermore, by taking the summation over all the possibilities of $\bfm y^{(q)}$ in~\eqref{eq_above}, we obtain
\begin{align}\label{eq_p(y^{(q)}|y)even}
\sum_{\bfm y^{(q)}}p(\bfm y^{(q)}|\bfm M=m)\sum_{\bfm y,p(\bfm y)\neq 0}p(\bfm y^{(q)}|\bfm y,\bfm M=m)&\leq \sum_{\bfm y^{(q)}}p(\bfm y^{(q)}|\bfm M=m)\prod_{k=1}^{\frac{q}{2}}(p_k+p_{-k})^{m_k}\nonumber\\
&= \sum_{m_1+\cdots +m_{\frac{q}{2}}=m} {m\choose m_1,\cdots, m_\frac{q}{2}}\prod_{l=1}^{\frac{q}{2}}(p_l+p_{-l})^{m_l}\prod_{k=1}^{\frac{q}{2}}(p_k+p_{-k})^{m_k}\nonumber\\
&= \left(\sum_{k=1}^{\frac{q}{2}}(p_k+p_{-k})^2\right)^m.
\end{align}
By substituting the result of~\eqref{eq_p(y^{(q)}|y)even} in~\eqref{eq_Hy^{(q)}-H-2Q}, we obtain
\begin{eqnarray}
H(\bfm Y^{(q)})-H(\bfm Y)&\geq &-\log\left(\sum_{k=1}^{\frac{q}{2}}(p_k+p_{-k})^2\right)\sum_{m}mp(m)\nonumber\\
&=&-E\{\bfm M\}\log\left(\sum_{k=1}^{\frac{q}{2}}(p_k+p_{-k})^2\right),
\end{eqnarray}
which concludes the proof.
\end{IEEEproof}

\begin{lem}\label{lem_HY^{(q)}Y|X}
In any binary input $q$-ary output channel with synchronization errors, for any input distribution and any even $q$, we have
\begin{equation}
H({\bfm Y^{(q)}}|\bfm X)\leq H(\bfm Y|\bfm X)+E\{\bfm M\}H(p_{-\frac{q}{1}},\cdots ,p_{-1},p_1,\cdots ,p_{\frac{q}{2}}).
\end{equation}
\end{lem}
\begin{IEEEproof}
The proof is similar to the proof of Lemma~\ref{lem_HY^{(q)}Y|Xodd}.
\end{IEEEproof}

\section{Proof of Theorem~\ref{thm-AWGN-quan}}\label{app_quan1}
We first compute $H_\Delta=H(p_{-M},\cdots,p_{-1},p_{1},\cdots,p_{M})+\log(\Delta)$ and $\sum_{m=1}^{M}\frac{1}{\Delta}\left(p_{m}+p_{-m}\right)^2$ for $M\to\infty$ and $\Delta\to 0$. Then by employing the result of Theorem~\ref{thm-synch-KQeven}, we prove the theorem.

For large $M$, we have $p_m\cong f(1-m\Delta)\Delta$ with the understanding that the approximation becomes exact as $\Delta\to 0$ where $f(x)=\frac{1}{\sqrt{2\pi}\sigma}e^{-\frac{x^2}{2\sigma^2}}$. Therefore, for $H_\Delta=H(p_{-M},\cdots,p_{-1},p_{1},\cdots,p_{M})+\log(\Delta)$, we can write
\begin{align}\label{eq_HDelta}
\lim_{M\to\infty,\Delta\to 0} H_\Delta&= \lim_{M\to\infty,\Delta\to 0} -\sum_{m=1}^M \bigg[f(1-m\Delta)\log(f(1-m\Delta))+f(1+m\Delta)\log(f(1+m\Delta))\bigg]\Delta \nonumber\\
&= \int_{0}^{\infty} \bigg[f(1-x)\left(\log(\sqrt{2\pi}\sigma)+\frac{(1-x)^2}{2\sigma^2}\log(e)\right)+f(1+x)\left(\log(\sqrt{2\pi}\sigma)+\frac{(1+x)^2}{2\sigma^2}\log(e)\right)\bigg]dx\nonumber\\
&= \int_{-\infty}^{\infty} f(1-x)\left(\log(\sqrt{2\pi}\sigma)+\frac{(1-x)^2}{2\sigma^2}\log(e)\right)dx\nonumber\\
&=\log(\sqrt{2\pi}\sigma)+\frac{\log(e)}{2}.
\end{align}

On the other hand, for $\sum_{m=1}^{M}\frac{1}{\Delta}\left(p_{m}+p_{-m}\right)^2$, by letting $M\to\infty$ and $\Delta\to 0$, we obtain
\begin{align}\label{eq_logp^22}
\lim_{M\to\infty,\Delta\to 0} \sum_{m=1}^{M}\frac{1}{\Delta}\left(p_{m}+p_{-m}\right)^2&= \lim_{M\to\infty,\Delta\to 0} \sum_{m=1}^{M}\left(f(1-m\Delta)+f(1+m\Delta)\right)^2\Delta\nonumber\\
&=\int_{0}^\infty \left(f(1-x)+f(1+x)\right)^2dx\nonumber\\
&=\frac{1}{\sqrt{2\pi}\sigma}\int_{0}^\infty \left(f(\sqrt{2}(1-x))+f(\sqrt{2}(1+x))+e^{-\frac{1}{\sigma^2}}f(\sqrt{2}x)\right)dx\nonumber\\
&=\frac{1}{2\sqrt{\pi}\sigma}(1+e^{-\frac{1}{\sigma^2}}).
\end{align}

Using the results of~\eqref{eq_HDelta} and \eqref{eq_logp^22}, we can write
\begin{align}
\lim_{M\to\infty,\Delta\to 0}\bigg(H(p_{-M},\cdots,p_{-1},p_{1},\cdots,p_{M})&+\log\left(\sum_{m=1}^M(p_m+p_{-m})^2\right)\bigg)\nonumber\\
&= \lim_{M\to\infty,\Delta\to 0}\left(H_\Delta+\log\left(\sum_{m=1}^M\frac{1}{\Delta}(p_m+p_{-m})^2\right) \right)\nonumber\\
&=\log\left(\sqrt{\frac{e}{2}}(1+e^{-\frac{1}{\sigma^2}})\right).
\end{align}
Finally, by substituting this result into~\eqref{eq_LB-KQeven}, the proof follows.
\color{black}
\bibliographystyle{ieeetran}
\bibliography{myrefs}
\end{document}